\documentclass[12pt,aps,prd,superscriptaddress,showpacs,longbibliography,floatfix,nofootinbib]{revtex4-1}

\usepackage[utf8]{inputenc}
\usepackage{slashed}
\pdfoutput=1

\usepackage{color}
\usepackage{graphicx}   
\usepackage{bm}
\usepackage{amsmath}
\usepackage{amsfonts}
\usepackage{hyperref}

\def\be{\begin{equation}}
\def\ee{\end{equation}}
\def\ba{\begin{eqnarray}}
\def\ea{\end{eqnarray}}

\begin{document}

\title{Life at the Landau Pole}

\author{Paul Romatschke}
\affiliation{Department of Physics, University of Colorado, Boulder, Colorado 80309, USA}
\affiliation{Center for Theory of Quantum Matter, University of Colorado, Boulder, Colorado 80309, USA}

\begin{abstract}
  If a quantum field theory has a Landau pole, the theory is usually called 'sick' and dismissed as a candidate for an interacting UV-complete theory. In a recent study on the interacting 4d O(N) model at large N, it was shown that at the Landau pole, observables remain well-defined and finite. In this work, I investigate both relevant and irrelevant deformations of the said model at the Landau pole, finding that physical observables remain unaffected. Apparently, the Landau pole in this theory is benign. As a phenomenological application, I compare the O(N) model to  QCD, by identifying $\Lambda_{\overline{\rm MS}}$ with the Landau pole in the O(N) model.
\end{abstract}

\maketitle

\section{Motivation}

In the early days of quantum field theory, Landau and collaborators studied quantum electrodynamics in perturbation theory \cite{abrikosov1954elimination}. They found that QED has positive $\beta$-function in perturbation theory, recognizing that this would lead to an uncontrolled growth of the theory's coupling constant as a function of energy. In modern notation, the QED running coupling to leading order in perturbation theory becomes
\be
\label{qed}
\alpha(\bar\mu)=\frac{1}{\frac{2}{3\pi}\ln \frac{\Lambda_{LP}}{\bar\mu}}\,,
\ee
where it is customary 
to fix $\Lambda_{LP}=m_e e^{\frac{3 \pi}{2\alpha_0}}$ with $m_e$ the electron mass and $\alpha_0\simeq \frac{1}{137}$.

Landau noted that besides the dependence of the fine-structure constant $\alpha$ on the momentum scale $\bar\mu$, the form (\ref{qed}) implied that the running coupling is diverging (has a pole) at a finite momentum scale $\bar\mu=\Lambda_{LP}$. Since the coupling diverges at this scale, it seems that one cannot meaningfully probe momentum scales $\bar\mu>\Lambda_{LP}$ in QED, so the theory does not have a well-defined continuum limit. It has even been suggested that Landau was so disturbed by this feature that he quit working on quantum field theory\footnote{I thank Bill Zajc for pointing out that this statement is not true, assuming the the publication dates of papers are chronologically matched to his actual work. However, Landau's view of QFT seems to have been fairly negative after 1954, e.g. in Ref.~\cite{landau1956quantum} the authors write ``\textit{This brings us to the conclusion that point interaction is impossible in pure electrodynamics.}'' and in Ref.~\cite{Kirzhnits:1978fy} Kirzhnits and Linde write about Ref.~\cite{landau1969collected} as ``\textit{..., which even led some authors to the conclusion that the hamiltonian quantum field theory ``is dead''...}''.}\footnote{See Ref.~\cite[chapter 8]{Ioffe:2012re} for a first-hand historical account of Landau's approach to QED.}.

Modern physics deals with the issue of the Landau pole through a mix of denial and shoulder shrugging. Denial adherents will rightly point out that (\ref{qed}) was derived in perturbation theory, requiring $\alpha\ll 1$, so that as a consequence (\ref{qed}) cannot be expected to correctly capture features such as the Landau pole, where by definition $\alpha\rightarrow \infty$. Shrugging adherents will (also rightly) point out that
\be
\label{qedpole}
\Lambda_{LP}=m_e e^{\frac{3 \pi}{2\alpha_0}}\simeq 10^{280}\, {\rm MeV}
\ee
puts the scale of the Landau pole beyond the Planck scale, so that in practice it is entirely pointless to understand QED in that regime anyway. (However, it should be noted that in the full Standard Model, $\Lambda_{LP}\simeq 10^{34}$ GeV is much lower than (\ref{qedpole}), but still extremely high \cite{Gockeler:1997dn}.) The prevailing dogma in both cases is that theories with a Landau pole should be viewed as 'UV-incomplete', or cut-off theories, that cannot be used to describe continuum physics.


In this work, I will entertain an entirely different perspective, namely that physical observables of a quantum field theory could be well-behaved even when the coupling diverges at the Landau pole, and beyond.

Unfortunately, I am unable to test my perspective in QED (yet), even though other groups have proposed similar ideas \cite{Kogut:1987cd,Kogut:1988sf}. Instead, I will focus on a quantum field theory which can be solved non-perturbatively in the limit of a large number of components, namely the O(N) model \cite{Romatschke:2022jqg}, using critically important input from $\mathcal{PT}$-symmetric field theory \cite{Ai:2022csx}. I will work in 3+1 dimensions where the O(N) model is known to possess a Landau pole in the large N limit.

In a quantum field theory, the renormalized coupling is not directly observable -- as is apparent through the fact that it will depend on the fictitious renormalization scale $\bar\mu$. For this reason, finite and well-defined physical observables at infinite renormalized coupling are certainly possible. Indeed, the idea that physical observables turn out to be finite even when the coupling diverges is well supported by several quantum field theory examples, such as ${\cal N}=4$ SYM in 3+1 dimensions \cite{Itzhaki:1998dd,Policastro:2001yc}, bosonic and fermionic large N field theories in 2+1 dimensions \cite{Romatschke:2019ybu,DeWolfe:2019etx,Romatschke:2021imm,Pinto:2020nip}, as well as non-relativistic fermions at a Feshbach resonance where experimental confirmation is available \cite{gurarie2007resonantly}.

In this work, I build upon and extend this idea: if physical observables remain finite when the renormalized coupling parameter diverges, maybe observables remain finite and well-defined when the renormalized coupling parameter becomes negative or even complex. After all, the renormalized coupling parameter  is not directly observable, so nothing should protect it from becoming complex as long as physical observables remain well-defined. Of course this type of idea cannot be tested in perturbation theory, which is inherently a weak-coupling expansion around non-interacting field theory. For this reason, I heavily employ large N expansion techniques (which do not rely on a small perturbative coupling in order to be applicable) as well as results from $\mathcal{PT}$-symmetric field theory (which allow calculations for negative or complex couplings via analytic continuation).

While this study is exploratory, I nevertheless hope that certain aspects merit further consideration when trying to interpret quantum field theory in four dimensions.

\section{Calculation -- The O(N) model in 3+1 dimensions at large N}

Let me first consider the case of the massless theory with quartic interaction, which is essentially a repeat of the calculation in Ref.~\cite{Romatschke:2022jqg}, but included here for completeness. The partition function for this theory is defined through the path integral
\be
Z(\lambda,\beta)=\int {\cal D}\phi e^{-S_E}\,,
\ee
with the Euclidean action
\be
\label{origaction}
S_E=\int d^3x \int_0^\beta d\tau \left[\frac{1}{2}\partial_\mu \vec{\phi}\cdot \partial_\mu \vec{\phi}+\frac{\lambda}{N} \left(\vec{\phi}\cdot \vec{\phi}\right)^{2}\right]\,,
\ee
where $\vec{\phi}=\left(\phi_1,\phi_2,\ldots,\phi_N\right)$ is an N-component scalar field and the theory is defined on the thermal cylinder with $\beta=\frac{1}{T}$ the inverse temperature. 

The partition function may be rewritten in a more convenient form by introducing two auxiliary fields $\sigma,\zeta$ with (cf. Ref.~\cite{Romatschke:2019rjk})
\be
\label{auxi}
1=\int{\cal D}\sigma \delta(\sigma-\vec{\phi}^2)=\int {\cal D}\sigma\int {\cal D}\zeta e^{i \int \zeta (\sigma-\vec{\phi^2})}\,.
\ee
The resulting path integral for $\sigma$ has quadratic action, such that $\sigma$ can be integrated out. One finds
\be
Z(\lambda,\beta)=\int {\cal D}\phi {\cal D}\zeta e^{-S_{ \rm eff}}\,,
\quad S_{\rm eff}=\int d^3x \int_0^\beta d\tau \left[\frac{1}{2}\vec{\phi}\left[-\Box+2 i \zeta\right]\vec{\phi}+N \frac{\zeta^2}{4 \lambda}\right]\,.
\ee
Separating the auxiliary field into zero modes and fluctuations $\zeta(x)=\frac{\zeta_0}{2}+\zeta^\prime(x)$, one can verify that the path integral over fluctuations does not contribute to leading order in large N to the partition function. 
Since $\zeta_0$ is a constant, the path integral over fields $\vec{\phi}$ is quadratic and can be done in closed form. One finds $Z(\lambda,\beta)=\int d\zeta_0 e^{N \beta V p(\sqrt{2 i \zeta_0})}$, with the pressure per component in dimensional regularization
\be
\label{bosonp}
p(m)=\frac{m^4}{16\lambda}+\frac{m^4}{64\pi^2}\left(\frac{1}{\varepsilon}+\ln\frac{\bar\mu^2 e^{\frac{3}{2}}}{m^2}\right)+\frac{m^2 T^2}{2\pi^2}\sum_{n=1}^\infty \frac{K_2(n \beta m)}{n^2}\,,
\ee
where $\beta V$ is the space-time volume, $\bar\mu$ is the $\overline{\rm MS}$ renormalization scale, $K_i(x)$ denotes modified Bessel functions of the second kind and I have rewritten $i \zeta_0=\frac{m^2}{2}$ to simplify the appearance.

\begin{figure}
  \includegraphics[width=.7\linewidth]{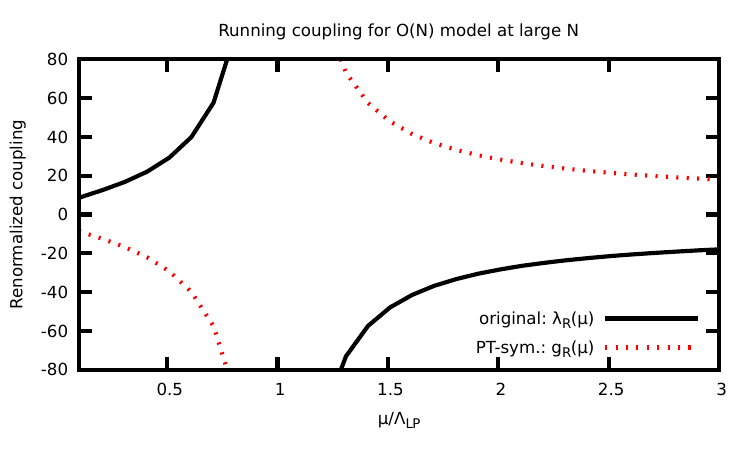} 
  \caption{Running coupling in the O(N) model in 3+1 dimensions. Shown are results for the coupling $\lambda_R$ in the original theory (\ref{prel2}) as well as for the coupling $g_R$ in the analytically continued theory ($\mathcal{PT}$-symmetric theory). Adapted from Ref.~\cite{Romatschke:2022jqg}.\label{fig1}}
  \end{figure}

The expression (\ref{bosonp}) is divergent in the continuum $\lim \varepsilon\rightarrow 0$. However, it may be non-perturbatively renormalized as
\be
\label{prebosonrun}
\frac{1}{\lambda}+\frac{1}{4\pi^2\varepsilon}=\frac{1}{\lambda_R(\bar\mu)}\,,
\ee
which is standard procedure for large N field theories \cite{Moshe:2003xn}. The resulting running coupling is given by
\be
\label{prel2}
\lambda_R(\bar\mu)=\frac{4\pi^2}{\ln \frac{\Lambda_{LP}^2}{\bar\mu^2}}\,,
\ee
which has a Landau pole at $\bar\mu=\Lambda_{LP}$, cf. Fig.~\ref{fig1}.

In order to make sense of the theory at the Landau pole, a procedure for analytically continuing the theory $\textit{beyond}$ the Landau pole is necessary. This procedure has been provided in the form of a conjecture in Ref.~\cite{Ai:2022csx} for so-called $\mathcal{PT}$-symmetric field theory\footnote{Note that the conjecture in Ref.~\cite{Ai:2022csx} has been formulated at zero temperature and checks exist only for $d=1$ (quantum mechanics). I thank W.~Ai for pointing this out to me.}. Naively continuing (\ref{prel2}) for $\bar\mu>\Lambda_{LP}$, the sign of $\lambda_R$ becomes negative. So in order to make sense of the O(N) model beyond the Landau pole, one is led to consider a theory where the  sign of the coupling is flipped:
\be
\label{relation}
\lambda\rightarrow -g+i 0^+\,,
\ee
where the small imaginary part has been included in order to be able to 'go around' the Landau pole. Following standard nomenclature \cite{Bender:1998ke}, the theory with flipped-sign coupling is referred to as $\mathcal{PT}$-symmetric field theory, and its partition function is denoted by $Z_{\mathcal{PT}}(g,\beta)$. Following Ref.~\cite{Ai:2022csx}, the analytic continuation of $Z(\lambda,\beta)$ is given by
\be
\ln Z_{\mathcal{PT}}(g,\beta)={\rm Re}\ln Z(\lambda=-g+i 0^+,\beta)\,.
\ee
To evaluate $Z_{\mathcal{PT}}(g,\beta)$, one may directly employ the pressure function (\ref{bosonp}), where now the sign of the coupling has been flipped. Regardless of the sign of the coupling, the expression for the pressure (\ref{bosonp}) is divergent in the 4d continuum $\lim \varepsilon\rightarrow 0$. The $\mathcal{PT}$-coupling renormalization, given by 
\be
\label{bosonrun}
\frac{1}{g}-\frac{1}{4\pi^2\varepsilon}=\frac{1}{g_R(\bar\mu)}\,,
\ee
differs from (\ref{prebosonrun}) by a sign, making the $\mathcal{PT}$-symmetric theory asymptotically free. Using the renormalized coupling $g_R(\bar\mu)$, one can express the renormalized pressure function as
\be
\label{p2}
p(m)=-\frac{m^4}{16g_R(\bar\mu)}+\frac{m^4}{64\pi^2}\ln\frac{\bar\mu^2 e^{\frac{3}{2}}}{m^2}+\frac{m^2 T^2}{2\pi^2}\sum_{n=1}^\infty \frac{K_2(n \beta m)}{n^2}\,.
\ee
The pressure, being a physical observable, cannot depend on the choice of renormalization scale $\bar\mu$, so $\frac{d p}{d\ln \bar\mu}=0$.  This fixes the running for the renormalized coupling $g_R(\bar\mu)$ and as a consequence the form of the running coupling itself as
\be
\label{l2}
g_R(\bar\mu)=\frac{4\pi^2}{\ln \frac{\bar\mu^2}{\Lambda_{LP}^2}}\,,
\ee
which is the same as (\ref{prel2}) up to a sign. Both running couplings are shown in Fig. ~\ref{fig1}, where it can be seen that they diverge at $\bar\mu=\Lambda_{LP}$. Whereas $\bar\mu=\Lambda_{LP}$ is the Landau pole in the O(N) model, when interpreted through ${\mathcal PT}$-symmetric field theory, $\Lambda_{LP}$ has all the trappings of an infrared scale parameter reminiscent of QCD (cf. section \ref{sec:QCD}).

For completeness, the ${\beta}$-function for the ${\cal PT}$-symmetric theory in $d=4-2\varepsilon$ dimensions with $\varepsilon\ll 1$ is explicitly calculated as
\be
\label{bgr}
\beta(g_R)=\frac{\partial g_R(\bar\mu)}{\partial \ln \bar\mu^2}=-g_R\left(\frac{g_R(\bar \mu)}{4 \pi^2}+\varepsilon\right)\,,
\ee
which is seen to be negative in $d=4$. Comparison with the perturbative (weak-coupling) result for $\beta$ \cite[Eq.(10.54)]{makeenko2002methods} shows that the large N $\beta$-function is only 1/9th of the N=1 weak-coupling result. The remainder of the complete perturbative contribution arises at next-to-leading order in the large N expansion.

From the form (\ref{bgr}) it is also easy to discuss the fix-point structure of the theory. In $d=4-2\varepsilon$ dimensions, the ${\cal PT}$-symmetric O(N) model possesses two fixed points, one at $g_R=0$ and the other at $g_R=-4 \pi^2 \varepsilon$. From the running coupling $g_R(\bar\mu)$ shown in Fig.~\ref{fig1}, it is clear that $g_R=-4\pi^2 \varepsilon$ corresponds to the IR fixed-point of the theory. Flipping the sign of both the coupling and $\varepsilon$, one finds that this fixed point is the same as the UV fixed point of the original theory specified by (\ref{origaction}) in $d=4+2\varepsilon$ dimensions. This is consistent with the known critical dimensions $4<d<6$ for which the O(N) model can be non-perturbatively renormalized \cite{Parisi:1975im}, see also \cite{Giombi:2019upv}.


%
Inserting the form (\ref{l2}) of the running coupling into (\ref{p2}), one obtains
\be
\label{sp2}
p(m)=\frac{m^4}{64\pi^2}\ln\frac{\Lambda_{LP}^2 e^{\frac{3}{2}}}{m^2}+\frac{m^2 T^2}{2\pi^2}\sum_{n=1}^\infty \frac{K_2(n \beta m)}{n^2}\,.
\ee

At this point it is worth to pause and consider the following observations:
\begin{itemize}
\item
  Any dependence of $p(m)$ on the unphysical renormalization scale $\bar\mu$ has dropped out
\item
  The expression (\ref{sp2}) is \textit{identical} to (\ref{p2}) evaluated at a the Landau pole, $\bar\mu=\Lambda_{LP}$
\item
  The pressure in the (flipped sign coupling and asymptotically free) $\mathcal{PT}$-symmetric theory (\ref{sp2})  is \textit{identical} to the the pressure in the original O(N) model (\ref{bosonp}), as can be seen by inserting Eqns.~(\ref{prebosonrun}), (\ref{prel2}) into (\ref{bosonp})
\item
  For generic values of $m$, the pressure at the Landau pole will be \textit{finite} rather than infinite.
\end{itemize}

As a consequence of these observations, the O(N) model and the $\mathcal{PT}$-symmetric O(N) model are identical at large N, even though from Fig.~\ref{fig1} one of these has a Landau pole and the other one is asymptotically free. For this reason I will refer to the scale $\bar\mu=\Lambda_{LP}$ as 'the Landau pole' also in the  $\mathcal{PT}$-symmetric theory.

One can calculate the pressure for any temperature $T$ by noting that the remaining single integral over $\zeta_0$ in the partition function is again dominated by the saddle points at large N, so that the physical pressure of the theory per component is given by (\ref{sp2}) with $m=\bar{m}$ the solution to
\be
\label{saddle}
0=\frac{d p(m)}{dm^2}=\frac{m^2}{32\pi^2}\ln \frac{\Lambda_{LP}^2e^1}{m^2}-\frac{mT}{4\pi^2}\sum_{n=1}^\infty \frac{K_1(n \beta m)}{n}\,.
\ee
If (\ref{saddle}) has more than one solution (as it generically does), then in the large N limit, the solution with the biggest real part of $p(\bar{m})$ will dominate over all others.

At zero temperature (a.k.a. the vacuum), a simple solution to (\ref{saddle}) is $\bar m=0$, which corresponds to the usual starting point for perturbative calculations. However, there is a second solution to (\ref{saddle}) located at $\bar m=\Lambda_{LP}\sqrt{e}$, which is usually dismissed as 'being too close to the Landau pole' \cite{Moshe:2003xn}. However, the physical pressure per component for this second solution is given by
\be
\label{pres2}
p(m=\Lambda_{LP}\sqrt{e},T=0)=\frac{\Lambda_{LP}^4 e^2}{128\pi^2}\,,
\ee
which is perfectly finite. In addition, since $p(m=\Lambda_{LP}\sqrt{e})>p(m=0)$, the solution (\ref{pres2}) is thermodynamically preferred over the perturbative vacuum\footnote{It is perhaps worth mentioning that some early studies of the O(N) model in 3+1 dimensions came to the conclusion that it contains tachyons or an instability to spontaneous generation of a large vacuum expectation value $\langle \vec{\phi}\rangle$ \cite{Coleman:1974jh,Linde:1976qh}. However, as pointed out already in Ref.~\cite{Abbott:1975bn}, none of these are expected to happen for the thermodynamically preferred phase at  large N.}.

\begin{figure}
  \includegraphics[width=.7\linewidth]{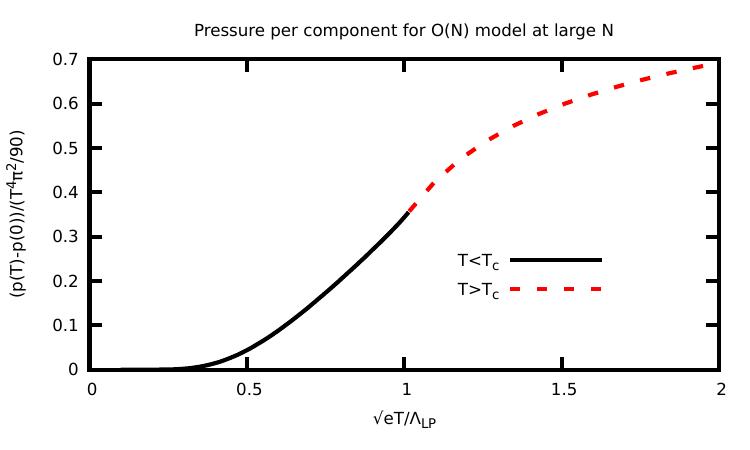}
  \caption{Pressure per component as a function of temperature for the O(N) model at large N, adapted from Ref.~\cite{Romatschke:2022jqg}. $T_c\simeq 0.616 \Lambda_{LP}$ denotes the location in temperature where the solution to (\ref{saddle}) for $m$ becomes complex. \label{fig1b}}
  \end{figure}

Despite the presence of the Landau pole, observables in the bosonic theory seem to make physical sense. For instance, one can calculate thermodynamic properties at finite temperature by tracking solutions to (\ref{saddle}) numerically, and evaluating $p(\bar{m})$. As discussed in Ref.~\cite{Romatschke:2022jqg}, at small temperature the numerical solution $\bar{m}$ of the thermodynamically preferred phase is continuously connected to $\bar{m}(T=0)=\sqrt{e}\Lambda_{LP}$. The numerical solution $\bar{m}$ becomes complex above a critical temperature $T=T_c\simeq \Lambda_{LP}/{\sqrt{e}}$, but using results from $\mathcal{PT}$-symmetric field theory \cite{Ai:2022csx}, the analytically continued pressure is continuous across $T=T_c$. A plot of the pressure as a function of temperature from Ref.~\cite{Romatschke:2022jqg} is reproduced in Fig.~\ref{fig1b}, and results for the entropy and specific heat can be found in Ref.~\cite{Romatschke:2022jqg}.

Besides the pressure, one may look at other observables. For instance, one can calculate the spectral function for the auxiliary field $\zeta^\prime$, finding a stable bound state in the low temperature phase \cite{Abbott:1975bn,Romatschke:2019gck}. This is \textit{different} from the situation in QED, where one encounters a tachyon (also sometimes called Landau's ghost) instead of a stable bound state \cite{Kirzhnits:1978fy}. This is the main reason why this work is concerned with the Landau pole in the O(N) model rather than QED.

\subsection{Adding Relevant Deformations}

One might worry that the results from the previous sections are an artifact of tuning away all relevant and irrelevant operators. For this reason, it is useful to consider repeating the analysis for the Euclidean action
\be
S_E=\int d^3x d\tau \left[\frac{1}{2}\partial_\mu \vec{\phi}\cdot \partial_\mu \vec{\phi}+\frac{1}{2}m_{\rm bare}^2\vec{\phi}^2-\frac{g}{N} \left(\vec{\phi}\cdot \vec{\phi}\right)^{2}\right]\,.
\ee
Introducing the auxiliary fields as before in (\ref{auxi}), one may again integrate out $\sigma$, and one finds in complete analogy with the previous section the ${\mathcal{PT}}$-symmetric pressure function
\be
\label{massp2}
p(m)=-\frac{(m^2-m_{\rm bare}^2)^2}{16g}+\frac{m^4}{64\pi^2}\left(\frac{1}{\varepsilon}+\ln\frac{\bar\mu^2 e^{\frac{3}{2}}}{m^2}\right)+\frac{m^2 T^2}{2\pi^2}\sum_{n=1}^\infty \frac{K_2(n \beta m)}{n^2}\,.
\ee

The explicit $\varepsilon\rightarrow 0$ divergence can again be taken care of by using the same non-perturbative renormalization as before (\ref{bosonrun}). This leads to
\be
\label{massp3}
p(m)=\frac{2 m^2 m_{\rm bare}^2-m_{\rm bare}^4}{16g}+\frac{m^4}{64\pi^2}\ln\frac{\Lambda_{LP}^2 e^{\frac{3}{2}}}{m^2}+\frac{m^2 T^2}{2\pi^2}\sum_{n=1}^\infty \frac{K_2(n \beta m)}{n^2}\,.
\ee

Since the bare coupling $\frac{1}{g}$ diverges as $\varepsilon\rightarrow 0$, there are residual divergences remaining in (\ref{massp3}). The first one of these can be taken care of by renormalizing the bare mass parameter as
\be
\frac{m_{\rm bare}^2}{g}=\frac{m_R^2(\bar\mu)}{g_R(\bar\mu)}\,,
\ee
The running of the renormalized mass $m_R(\bar\mu)$ is again fixed by requiring that $\frac{d p(m)}{d\ln \bar\mu}=0$, which leads to
\be
\label{massrendef}
m_R^2(\bar\mu)=\frac{\rm const}{\ln \frac{\bar\mu^2}{\Lambda_{LP}^2}}\,, \quad {\rm or}\quad
\frac{m_R^2(\bar\mu)}{g_R(\bar\mu)}=m_0^2\,,
\ee
with constant and finite mass scale $m_0$. This leads to
\be
\label{massp4}
p(m)=\frac{m^2 m_0^2}{8}-\frac{m_{\rm bare}^4}{16g}+\frac{m^4}{64\pi^2}\ln\frac{\Lambda_{LP}^2 e^{\frac{3}{2}}}{m^2}+\frac{m^2 T^2}{2\pi^2}\sum_{n=1}^\infty \frac{K_2(n \beta m)}{n^2}\,.
\ee
For the remaining term, note that in the limit $\varepsilon\rightarrow 0$
\be
\frac{m_{\rm bare}^4}{g}=m_0^2 m_{\rm bare}^2=m_0^4 g=\frac{m_0^4}{\frac{1}{g_R(\bar\mu)}+\frac{1}{4\pi^2\varepsilon}}\rightarrow 0\,, 
\ee
so that the pressure function becomes
\be
\label{massp5}
p(m)=\frac{m^2 m_0^2}{8}+\frac{m^4}{64\pi^2}\ln\frac{\Lambda_{LP}^2 e^{\frac{3}{2}}}{m^2}+\frac{m^2 T^2}{2\pi^2}\sum_{n=1}^\infty \frac{K_2(n \beta m)}{n^2}\,.
\ee

For small values of $m_0$, the properties of this theory are close to the unmodified version considered in the previous section. The second-order phase transition at finite temperature persists, but is pushed to higher values of $\frac{T_c}{\Lambda_{LP}}$.


\subsection{Adding Irrelevant Deformations}

Now let us consider what happens when adding irrelevant operators to the theory. In this case, I study
\be
S_E=\int d^3x  d\tau \left[\frac{1}{2}\partial_\mu \vec{\phi}\cdot \partial_\mu \vec{\phi}-\frac{g}{N} \left(\vec{\phi}\cdot \vec{\phi}\right)^{2}+\frac{\alpha}{N^2}\left(\vec{\phi}\cdot \vec{\phi}\right)^{3}\right]\,,
\ee
where $\alpha$ is the bare sextic coupling parameter.  Introducing the auxiliary fields as before in (\ref{auxi}), it is possible, but not very enlightening, to integrate out $\sigma$ exactly. Instead, in the large N limit it is again permissible to replace $\sigma(x)$ by just its global zero mode $\sigma_0$, so that
\be
Z_{\mathcal PT}(g,\beta)=\int d\sigma_0 d\zeta_0 e^{N \beta V p(m=\sqrt{i \zeta_0}, \sigma_0)}\,,
\ee
where
\be
\label{sxp2}
p(m,\sigma_0)=\frac{g \sigma_0^2}{N^2}-\frac{\alpha \sigma_0^3}{N^3}+\frac{\sigma_0 m^2}{2 N}+\frac{m^4}{64\pi^2}\left(\frac{1}{\varepsilon}+\ln\frac{\bar\mu^2 e^{\frac{3}{2}}}{m^2}\right)+\frac{m^2 T^2}{2\pi^2}\sum_{n=1}^\infty \frac{K_2(n \beta m)}{n^2}\,.
\ee
At large N, the integral over $\sigma_0$ is done with the saddle point method, with two saddles located at
\be
\label{sxsigma}
\sigma_0^{(1,2)}=\frac{g N}{3\alpha}\left(1\pm \sqrt{1+\frac{3 \alpha m^2}{2g^2}}\right)\,.
\ee
For small $\frac{3 \alpha m^2}{2g^2}$ (justified below), one can expand the square root in this expression to obtain
\ba
\frac{\sigma_0^{(1)}}{N}&=&-\frac{m^2}{4g}+\frac{3\alpha m^4}{32 g^3}-\frac{9 \alpha^2 m^6}{128 g^5}+\sum_{n=3}^\infty{\cal O}\left(\frac{\alpha^{2n}}{g^{2n+1}}\right)\,,\nonumber\\
\frac{\sigma_0^{(2)}}{N}&=&\frac{2 g}{3\alpha}+\frac{m^2}{4g}-\frac{3\alpha m^4}{32 g^3}+\frac{9 \alpha^2 m^6}{128 g^5}+\sum_{n=3}^\infty{\cal O}\left(\frac{\alpha^{2n}}{g^{2n+1}}\right)\,,
\ea
for the two solutions. Inserting $\sigma_0^{(1)}$ into (\ref{sxp2}), one finds
\be
\label{sxp3one}
p(m)=-\frac{m^4}{16 g}+\frac{\alpha m^6}{64 g^3}+\frac{m^4}{64\pi^2}\left(\frac{1}{\varepsilon}+\ln\frac{\bar\mu^2 e^{\frac{3}{2}}}{m^2}\right)+\frac{m^2 T^2}{2\pi^2}\sum_{n=1}^\infty \frac{K_2(n \beta m)}{n^2}+\sum_{n=1}^\infty{\cal O}\left(\frac{m^{2n+6}\alpha^{2n}}{g^{2n+3}}\right)\,.
\ee
Renormalizing the coupling $g$ as in (\ref{bosonrun}), leads to
\be
\label{sxp4one}
p(m)=-\frac{m^4}{16 g_R(\bar\mu)}+\frac{\alpha m^6}{64 g^3}+\frac{m^4}{64\pi^2}\ln\frac{\bar\mu^2 e^{\frac{3}{2}}}{m^2}+\frac{m^2 T^2}{2\pi^2}\sum_{n=1}^\infty \frac{K_2(n \beta m)}{n^2}+\sum_{n=1}^\infty{\cal O}\left(\frac{m^{2n+6}\alpha^{2n}}{g^{2n+3}}\right)\,,
\ee
but this implies that $\frac{\alpha}{g^3}$ is divergent. Thus the bare sextic coupling parameter also needs to be renormalized as
\be
\frac{\alpha}{g^3}=\frac{\alpha_R(\bar\mu)}{g^3_R(\bar\mu)}\,.
\ee
This leaves the whole tower of additional terms $\frac{\alpha^{2n}}{g^{2n+3}}$, $n\geq 1$ that are potentially divergent. However, one finds that in the $\varepsilon\rightarrow 0$ limit
\be
\label{reason}
\frac{\alpha^{2n}}{g^{2n+3}}=\frac{\alpha_R^{2n}(\bar\mu)}{g_R^{6n}(\bar\mu)} \frac{1}{g^{3-4n}}=\frac{\alpha_R^{2n}(\bar\mu)}{g_R^{6n}(\bar\mu)} \frac{1}{\left(\frac{1}{g_R(\bar\mu)}+\frac{1}{4\pi^2 \varepsilon}\right)^{4n-3}}\rightarrow 0\,,
\ee
because $n\geq 1$. Therefore, none of these terms contribute, and one is left with
\be
\label{sxp5one}
p(m)=\frac{m^6}{M^2}+\frac{m^4}{64\pi^2}\ln\frac{\Lambda_{LP}^2 e^{\frac{3}{2}}}{m^2}+\frac{m^2 T^2}{2\pi^2}\sum_{n=1}^\infty \frac{K_2(n \beta m)}{n^2}\,,
\ee
where I have used the renormalization group invariance of the pressure to express
\be
\label{alpharen}
\frac{\alpha_R(\bar\mu)}{64 g_R^3(\bar\mu)}=\frac{1}{M^2}\,,
\ee
with constant mass scale $M$. One observes that for the same reason as (\ref{reason}), expanding the square root in (\ref{sxsigma}) is justified for $\sigma_0^{(1)}$.

For the second solution $\sigma_0=\sigma_0^{(2)}$, (\ref{sxp2}) becomes
\be
\label{sxp3two}
p(m)=\frac{4 g^3}{27 \alpha^2}+\frac{g m^2}{3\alpha}+\frac{m^4}{16 g}-\frac{\alpha m^6}{64 g^3}+\frac{m^4}{64\pi^2}\left(\frac{1}{\varepsilon}+\ln\frac{\bar\mu^2 e^{\frac{3}{2}}}{m^2}\right)+\frac{m^2 T^2}{2\pi^2}\sum_{n=1}^\infty \frac{K_2(n \beta m)}{n^2}+\ldots 
\ee
The explicit $\frac{1}{\varepsilon}$ divergence can by renormalizing the coupling g, but the sign of the counterterm must be flipped (and as a consequence, so must  the sign of the running coupling). One obtains
\be
\label{sxp4two}
p(m)=\frac{4 g^3}{27 \alpha^2}+\frac{g m^2}{3\alpha}+\frac{m^4}{16 g_R(\bar\mu)}-\frac{\alpha m^6}{64 g^3}+\frac{m^4}{64\pi^2}\ln\frac{\bar\mu^2 e^{\frac{3}{2}}}{m^2}+\frac{m^2 T^2}{2\pi^2}\sum_{n=1}^\infty \frac{K_2(n \beta m)}{n^2}+\ldots 
\ee
Renormalizing the sextic coupling $\alpha$ as in (\ref{alpharen}), the square-root expansion in (\ref{sxsigma}) is justified also for $\sigma_0^{(2)}$. Similar to (\ref{reason}), terms with positive powers of the bare coupling $g$ in the numerator vanish, so that one finds
\be
\label{sxp5two}
p(m)=\frac{M^4}{27\,648\,g^3}-\frac{m^6}{M^2}+\frac{m^4}{64\pi^2}\ln\frac{\Lambda_{LP}^2 e^{\frac{3}{2}}}{m^2}+\frac{m^2 T^2}{2\pi^2}\sum_{n=1}^\infty \frac{K_2(n \beta m)}{n^2}\,
\ee
This expression still has a divergent term $\propto \frac{1}{g^3}$, but this term is independent from $m$. For this reason, this last divergence can be canceled by a vacuum pressure-counterterm in the Lagrangian, leading to
\be
\label{sxp6two}
p(m)=-\frac{m^6}{M^2}+\frac{m^4}{64\pi^2}\ln\frac{\Lambda_{LP}^2 e^{\frac{3}{2}}}{m^2}+\frac{m^2 T^2}{2\pi^2}\sum_{n=1}^\infty \frac{K_2(n \beta m)}{n^2}\,.
\ee
Inspecting (\ref{sxp5one}) and (\ref{sxp6two}), one finds that the two saddle point solutions for $\sigma_0$ give rise to the same form for the pressure function, except for the sign of the $m^6$ term, which can be attributed to the fact that the sign of $g_R(\bar\mu)$ is flipped for the solution $\sigma_0^{(1)}$. 

The final integral over $\zeta_0$ is done by finding the saddle point solution
\be
\label{su1}
0=\frac{dp}{dm^2}=\pm \frac{3m^4}{M^2}+\frac{m^2}{32\pi^2}\ln \frac{\Lambda_{LP}^2e^1}{m^2}-\frac{m T}{4\pi^2}\sum_{n=1}^\infty \frac{K_1(n \beta m)}{n}\,,
\ee
where $\pm$ corresponds to the solutions $\sigma_{0}^{(1,2)}$, respectively. At zero temperature, where the contribution from the modified Bessel function vanishes, there is a different number of solutions depending on the sign in (\ref{su1}) and magnitude of $M^2$. For positive sign (corresponding to solution $\sigma_0^{(1)}$ above), and large $M^2$, there are three solutions: $\bar m=0$, $\bar m\simeq \Lambda_{LP}\sqrt{e}$ and $\bar m \propto M$ up to logarithmic corrections. Of these, the saddle with the largest pressure (lowest free energy) is $\bar m\simeq  \Lambda_{LP}\sqrt{e}$, hence this is the dominant saddle point at large N. One thus again recovers the solution (\ref{pres2}). As $M^2$ decreased, the situation remains qualitatively the same until $M\lesssim 84 \Lambda_{LP}$, at which point solutions (except for $\bar m=0$) become complex-valued. However, close to the Landau pole where $g_R(\bar\mu=\Lambda_{LP})\rightarrow \infty$, (\ref{alpharen}) suggests that $M^2\rightarrow \infty$, so I will not consider small values of $M$ in the following.

For the negative sign in (\ref{su1}) and large $M^2$, there are two solutions $\bar m=0$ and $\bar m \simeq \Lambda_{LP}\sqrt{e}$, where again the second solution is thermodynamically preferred. Therefore, one also recovers  the unmodified theory solution (\ref{pres2}) for the second solution $\sigma_0^{(2)}$.

\subsection{Observables and Running Coupling}
\label{sec:obsis}

As mentioned above, the running coupling of a quantum field theory is not directly observable, because it is a renormalization-scheme dependent quantity. However, observables in a quantum field theory can depend in a non-trivial manner on the coupling parameter of the theory.

For instance, it is possible to calculate scattering amplitudes in the O(N) model, which are closely related to observables such as cross-sections. It turns out that the Euclidean momentum s-wave scattering amplitude to leading order in large N is given by the propagator of the auxiliary field $\zeta^\prime$ introduces in (\ref{auxi}), which for the ${\cal PT}$-symmetric theory becomes \cite{Romatschke:2022jqg}
\be
{\cal M}(k)=-D(k)=\frac{32\pi^2}{N}\frac{1}{\frac{4\pi^2}{g_R(\bar\mu)}-\ln \frac{\bar\mu^2 e^2}{m^2}+2 \sqrt{\frac{k^2+4 m^2}{k^2}}{\rm atanh}\sqrt{\frac{k^2}{k^2+4m^2}}}\,.
\ee
I will use ${\cal M}(k)$ as an example of an observable quantity.

In weak coupling perturbation theory, one can expand ${\cal M}(k)$ for $g_R\ll 1$, so
\be
{\cal M}(k)\sim \frac{8}{N} g_R(\bar\mu)+{\cal O}(g_R^2)\,.
\ee
From a perturbative QFT point of view, one can therefore \textit{define an ``observable'' or ``effective'' coupling} as
\be
\label{geffdeff}
g_{R,eff}(\bar\mu\simeq k)\equiv \frac{N}{8}{\cal M}(k)\,,
\ee
where the identification $\bar\mu \simeq k$ is expected to hold only approximately. This is a reasonable thing to do as long as the resulting $g_{R,eff}$ is small enough in order for the observable ${\cal M}$ to be well approximated by perturbation theory.

\begin{figure}[t]
  \centering
  \includegraphics[width=.7\linewidth]{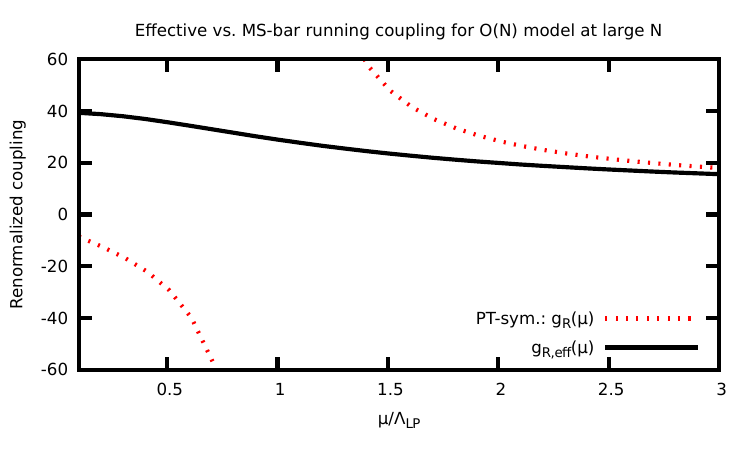}
  \caption{``Effective'' running coupling (\ref{geffdeff2}) vs. $\overline{\rm MS}$ coupling (\ref{l2}) for the ${\cal PT}$-symmetric O(N) model at large N. The effective coupling $g_{R,eff}$ is defined from the ``physical'' scattering amplitude ${\cal M}$ and is positive and finite for all values of $\bar\mu$. See text for details. \label{fig2b}}
  \end{figure}

However, since ${\cal M}(k)$ is an observable, it cannot depend on $\bar\mu$, and indeed it does not. Using the explicit form of $g_R(\bar\mu)$ from (\ref{l2}), one finds
\be
{\cal M}(k)=\frac{32\pi^2}{N}\frac{1}{-\ln \frac{\Lambda_{LP}^2 e^2}{m^2}+2 \sqrt{\frac{k^2+4 m^2}{k^2}}{\rm atanh}\sqrt{\frac{k^2}{k^2+4m^2}}}\,.
\ee
This result is fully non-perturbative, and can be further simplified by noting that to this order in the 1/N expansion, the value of $m$ is given by its saddle point value $m=\bar m=\sqrt{e}\Lambda_{LP}$, such that
\be
{\cal M}(k)=\frac{32\pi^2}{N}\frac{1}{-1+2 \sqrt{\frac{k^2+4 \bar m^2}{k^2}}{\rm atanh}\sqrt{\frac{k^2}{k^2+4\bar m^2}}}\,.
\ee
For sufficiently large Euclidean momenta $k\gg \bar m$ where ${\cal M}(k)$ becomes small, this leads to
\be
{\cal M}(k\gg \bar m)=\frac{8}{N}\frac{4\pi^2}{\ln\frac{k^2}{\Lambda_{LP}^2 e^2}}\,.
\ee
Comparing this to (\ref{geffdeff}) and (\ref{l2}) informs the refined definition for the effective coupling
\be
\label{geffdeff2}
g_{R,eff}(\bar\mu=k)\equiv \frac{N}{8}{\cal M}(k e^{1})\,.
\ee

I am now in a position to compare the ``physical'' definition of the coupling $g_{R,eff}(\bar\mu=k)$ to the running coupling of the theory (\ref{l2}), which I do in Fig.~\ref{fig2b}. From this figure, it can be seen that the effective coupling $g_{R,eff}(\bar\mu)$ is positive and finite for all energy scales $\bar\mu$. Secondly, one finds that the effective coupling $g_{R,eff}(\bar\mu)$ reaches a finite value in the infrared:
\be
\lim_{\bar\mu \rightarrow 0} g_{R,eff}(\bar\mu)=4\pi^2\,.
\ee

It is trivial to see that these properties of the effective coupling $g_{R,eff}$ are a direct consequence of the definition (\ref{geffdeff2}) in terms of a ``physical'' quantity (the scattering amplitude). In particular, $g_{R,eff}$ does not exhibit any remarkable feature near $\bar\mu=\Lambda_{LP}$ despite the fact that the O(N) model possesses a Landau pole at this scale.

This makes sense: since the running coupling itself is not an observable, one can always find some ``physical'' definition of a coupling that agrees with perturbation theory whenever the coupling is small, and has any set of properties one would want such as positiveness and finiteness in the IR. 

\subsection{Conclusions}

The O(N) model in 3+1 dimensions has a Landau pole at large N. Physical observables at the Landau pole remain finite and well-behaved. This feature does not change when adding either relevant operators (e.g. mass terms) or irrelevant operators (e.g. sextic interactions) to the theory. I therefore conclude that for the O(N) model in 3+1 dimensions at large N, the Landau pole is a harmless feature of the theory, and not a sign that the theory itself is 'sick'. Most importantly, using the analytic continuation provided by $\mathcal{PT}$-symmetric field theory, the O(N) model does not need to be treated as a cut-off theory. It is UV complete, and asymptotically free, despite (or perhaps because of ) the negative coupling constant.

\begin{figure}[t]
  \centering
  \includegraphics[width=.7\linewidth]{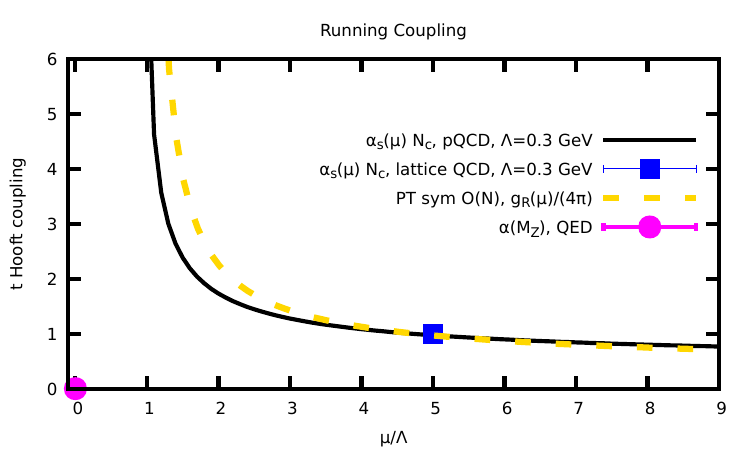}
  \caption{Running coupling in QCD (with $\Lambda\equiv \Lambda_{\overline{\rm MS}}=0.3$ GeV) and large N ${\mathcal{PT}}$-symmetric O(N) model.
In order to have a fair comparison, I'm converting the QCD running coupling $\alpha_s(\bar\mu)$ to the 't Hooft coupling by multiplying with a factor of $N_c=3$, and I am dividing the $\mathcal{PT}$-symmetric O(N) model coupling by a factor of $4\pi$. Lattice QCD point at $\bar\mu=1.5$ GeV is from Ref.~\cite{Bazavov:2014soa}, QED point at the Z-pole mass $\bar\mu=M_Z$ is from Ref.~\cite{ParticleDataGroup:2022pth}. See text for details. \label{fig3}}
  \end{figure}

An important omission in the present study is the question about $\frac{1}{N}$ corrections -- will they destroy the features of the large N limit? While this question is without doubt important, I cannot resist pointing out that (almost) the entirety of holography is built upon the strict large N limit of field theory, without systematic discussion of $\frac{1}{N}$ corrections \cite{Maldacena:1997re}. Yet even without systematic understanding of $\frac{1}{N}$ terms, holography has indisputable been useful in building our understanding of quantum field theory, so maybe a similar attitude could be extended to the large N limit of the O(N) model.

\section{A Landau pole in QCD?}

\label{sec:QCD}

Let me conclude the discussion by advancing a provocative idea: a Landau pole in Quantum Chromodynamics. According to lore, QCD does not have a Landau pole, so it seems that it should not be discussed here. But then again, there is no hard evidence against a Landau pole in QCD, so I would argue that the idea should be considered\footnote{See section \ref{sec:evi} for more discussion on this point.}. QCD does have asymptotic freedom, and since the $\mathcal{PT}$-symmetric O(N) model at large N shares this property, one can nevertheless ask how different or similar these theories are.

It would seem more appropriate to attempt a comparison to QED rather than QCD, but the experimental determination of the QED fine-structure constant does not extend much beyond the Z-pole mass $\bar\mu=M_Z\simeq 91$ GeV. Since the QED Landau pole (\ref{qedpole}) is at vastly higher energy, the resulting QED information (shown in Fig.~\ref{fig3}) is not particularly illuminating.

\begin{figure}
  \centering
  \includegraphics[width=.7\linewidth]{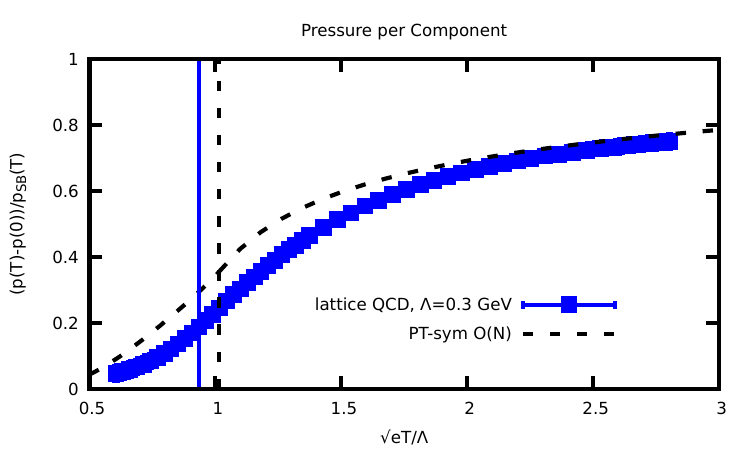}
  \caption{Pressure as a function of temperature for QCD (with $\Lambda\equiv \Lambda_{\overline{\rm MS}}=0.3$ GeV) and large N ${\mathcal{PT}}$-symmetric O(N) model. The QCD pressure is from lattice QCD with $N_f=2+1$ flavors of quarks with physical masses \cite{Borsanyi:2013bia}, with a cross-over temperature reported as $T_c=170(4)(3)$ MeV \cite{Aoki:2009sc} (full vertical line). Massless O(N) model results with a second-order phase transition located at $T_c\simeq 0.616 \Lambda_{LP}$ (dashed vertical line) are adapted from \cite{Romatschke:2022jqg}. See text for details. \label{fig4}}
  \end{figure}

So instead of QED, in Fig.~\ref{fig3} I compare the running coupling from QCD to that in the $\mathcal{PT}$-symmetric O(N) model.  For the QCD running coupling, I am using the 3-loop perturbative QCD expression resulting from numerically integrating
\be
\frac{\partial a_s}{\partial \ln \bar\mu^2}=-\beta_0 a_s^2-\beta_1 a_s^3-\beta_2 a_s^4+\ldots\,,
\ee
where $a_s=\frac{\alpha_s(\bar\mu)}{4\pi}$ and $\beta_0=11-\frac{2}{3}N_f$, $\beta_1=102-\frac{38}{3} N_f$, $\beta_2=\frac{2857}{2}-\frac{5033}{18}N_f+\frac{325}{54}N_f^2$ and I am taking $N_f=5$ \cite{vanRitbergen:1997va,ParticleDataGroup:2022pth}. As can be seen from Fig.~\ref{fig3}, the QCD running coupling obtained from perturbation theory becomes very large at small scales $\bar\mu$. In fact, one finds that the perturbative solution for $\alpha_s(\bar\mu)$ thus obtain diverges for a particular value of $\bar\mu=\Lambda_{\overline{\rm MS}}\simeq 0.3$ GeV. This value is consistent with similar values reported by other methods \cite{Proceedings:2019pra}, and indeed the running coupling $\alpha_s(\bar\mu)$ is consistent with calculations from lattice QCD \cite{Bazavov:2014soa} at values as low as $\bar\mu\simeq 5 \Lambda_{\overline{\rm MS}}$ (also shown in Fig.~\ref{fig3}).

In my opinion, Fig.~\ref{fig3} indicates a certain qualitative similarity between QCD and the O(N) model. Pushing the similarity further, this would lead to the interpretation that $\alpha_s(\bar\mu)$ actually does diverge at a finite momentum scale $\Lambda_{\overline{\rm MS}}$, and that deep in the infrared $\alpha_s$  should be analytically continued to negative (or complex) values.

\begin{figure}
  \centering
  \includegraphics[width=.45\linewidth]{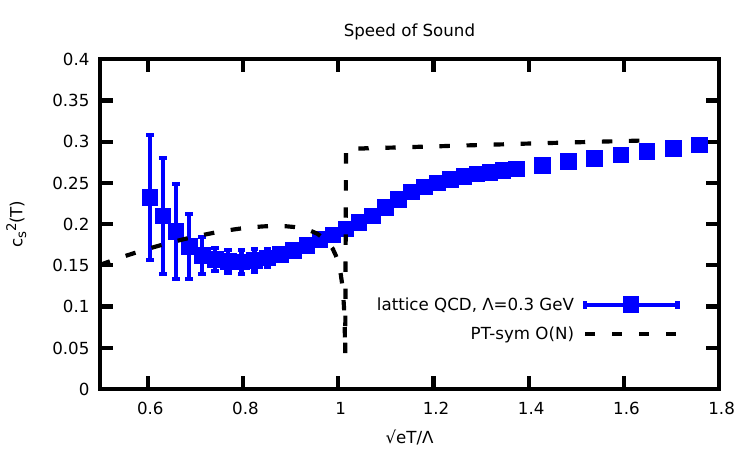}
  \hfill
  \includegraphics[width=.45\linewidth]{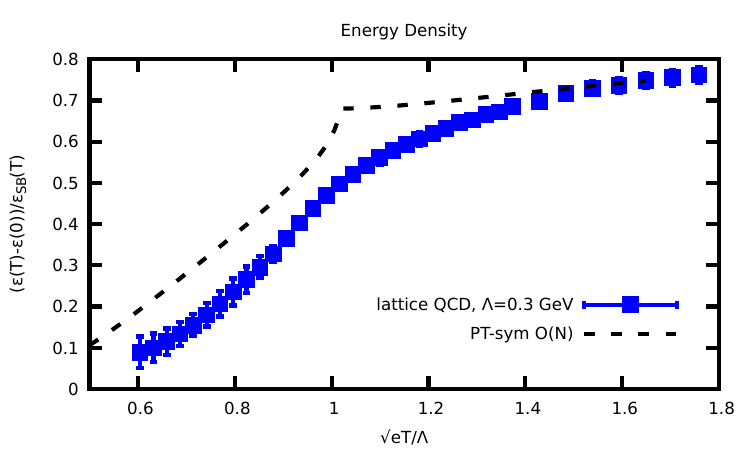}
  \caption{Speed of sound squared $c_s^2$ (left panel) and energy density $\epsilon(T)$ (right panel) as a function of temperature from lattice QCD \cite{Borsanyi:2013bia} (with $\Lambda\equiv \Lambda_{\overline{\rm MS}}=0.3$ GeV) and large N ${\mathcal{PT}}$-symmetric O(N) model. Unlike the pressure, both $c_s^2$ and $\epsilon(T)$ clearly show the second order phase transition in the O(N) model as compared to the analytic cross-over in QCD. See text for details. \label{fig5}}
  \end{figure}

It is well-known that QCD becomes confining in the infrared, and that physical observables are well-behaved and finite for all momentum scales. In particular, thermodynamic quantities such as the pressure are continuous as a function of temperature for QCD, with a broad analytic crossover from confined to quark-gluon plasma phase around $T_c\simeq 170$ MeV \cite{Aoki:2009sc}. Normalizing the pressure by the Stefan-Boltzmann pressure $p_{SB}(T)=\frac{\pi^2 T^4}{90}\left(2 (N_c^2-1)+\frac{7}{2} N_c N_f\right)$ with $N_c=N_f=3$ for QCD and $p_{SB}(T)=\frac{\pi^2 T^4 N}{90}$ for the O(N) model, a comparison is shown in Fig.~\ref{fig4}.

Similar to the case of the running coupling shown in Fig.~\ref{fig3}, thermodynamic properties for QCD and the O(N) model shown in Fig.~\ref{fig4} seem to have a certain qualitative similarity when expressed in units of $\Lambda_{\overline{\rm MS}}$. Upon closer inspection, however, it becomes clear that there are certain qualitative differences between QCD and the O(N) model, as evident when considering the energy density and speed of sound for both theories shown in Fig.~\ref{fig5}. The parameters relevant for the equation of state shown in this figure clearly exhibit the second-order phase transition in the O(N) model, which is absent in real-world QCD.

This was to be expected: the O(N) model simply is not the same as QCD, despite sharing the property of asymptotic freedom. At best, the O(N) model may serve as a crude approximation of QCD, similar to how ${\cal N}=4$ SYM has been used as a model for QCD \cite{Itzhaki:1998dd,Policastro:2001yc}. Nevertheless, given the successes of applying results from ${\cal N}=4$ SYM to QCD phenomenology \cite{Casalderrey-Solana:2011dxg}, it is possible that the O(N) model can serve as a useful phenomenological model, one which possesses a property that ${\cal N}=4$ SYM simply does not have: asymptotic freedom.

\subsection{On the evidence against a Landau pole in QCD}
\label{sec:evi}

There are multiple claims in the literature that in QCD, the Landau pole in the running coupling is cured by non-perturbative effects, see e.g. Refs.\cite{Antonov:2007mx,Deur:2016tte,Stevenson:2022gcv}. While some of these claims do seem not withstand rigorous scrutiny \cite{Stevenson:2023isz}, others are based on careful studies of non-perturbative results in QCD, e.g. in Refs. \cite{Dokshitzer:1998pt,Shoshi:2002rd}.

As a representative example, let me consider the observation that certain quantities such as the gluon polarization tensor appear to ``freeze'' in the infrared for QCD \cite{Antonov:2007mx}. This is analogous to the behavior of the scattering amplitude discussed for the O(N) model in section \ref{sec:obsis}. As in (\ref{geffdeff2}), the authors of Ref.~\cite{Antonov:2007mx} define an ``effective'' running coupling $\alpha_{s,eff}(\bar\mu)$ for QCD and show that it is well behaved and finite even in the zero momentum limit $\bar\mu\rightarrow 0$.

From the O(N) model example discussed in section \ref{sec:obsis}, it is clear that the approach taken in Ref.~\cite{Antonov:2007mx} is perfectly reasonable and useful as a practical definition for $\alpha_s$. However, from the O(N) model discussion it should also be clear that the observed infrared ``freezing'' of the coupling is entirely unrelated to the absence of a Landau pole. In fact, the O(N) model example shows that infrared freezing for an effective running coupling can occur in a theory with a Landau pole. To reiterate the point made in section \ref{sec:obsis}, the QCD running coupling is not an observable, so it is always possible to find some ``physical'' definition for $\alpha_s$ that agrees with perturbative QCD in the UV and has any set of properties in the infrared. While certain authors attribute great importance to the apparent ``infrared fixed point'' property based on the IR freezing of some suitably defined $\alpha_{s,eff}$, I personally do not believe that it helps in our understanding of QCD.

Of course, it is important to state that the similarities between the O(N) model, which possesses a Landau pole, and QCD, do not constitute evidence for the presence of a Landau pole in QCD. It could simply be that observables in a theory with a ``benign'' Landau pole such as the O(N) model are similar to a theory without a Landau pole.

To summarize, while there is no guarantee that a Landau pole exists for QCD, I am not aware of any hard evidence against one, either.

\section{Summary}

In this work, I have considered the O(N) model in 3+1 dimensions at large N, which has a Landau pole. Using technology borrowed from $\mathcal{PT}$-symmetric field theories, I have extended the theory beyond the Landau pole, and I have found that adding relevant and irrelevant operators do not qualitatively change the behavior of the theory close to the Landau pole. Physical observables in the  O(N) model are finite and well-behaved at and close to the Landau pole.

I take this to constitute evidence that at least at large N, the Landau pole in the O(N) model is harmless, and the theory does constitute a UV-complete interacting and asymptotically free theory.

Moreover, I compared results from the large N O(N) model for the running coupling and the finite temperature pressure to QCD, finding qualitative similarities between these two theories. Both the O(N) model and QCD are asymptotically free theories, but there is a key difference between these two theories. For QCD, the high energy regime possesses an intuitive interpretation as a weakly coupled equilibrium quantum field theory with well-defined particle states (quarks and gluons), whereas the infrared limit of QCD is notoriously difficult to formulate as an equilibrium continuum quantum field theory. By contrast, for the O(N) model it is the \textit{low energy regime} which lends itself to an intuitive equilibrium quantum field theory formulation (with the vector and scalar bound state the perturbative degrees of freedom), whereas the \textit{high energy region} is difficult to interpret. Coming full circle, while the respective ``perturbative'' and ``non-perturbative'' regimes are different for QCD and the O(N) model, it is curious to note that both theories contain both phases, and that the Landau pole acts as a the phase boundary for the O(N) model.

Based on these observations, my interpretation is as follows: Landau poles are common features for quantum field theories in 3+1 dimensions, since the O(N) model at large N, QED, and even QCD possess diverging coupling constants at a finite momentum scale $\bar\mu=\Lambda$. For two of these theories (O(N) model and QCD), we know that nothing 'bad' happens at this scale. Instead, $\bar\mu=\Lambda$ merely marks the scale at which the O(N) model and QCD seem to transition from a low temperature phase to a high temperature phase.

Perhaps it would be time to critically reassess the current dogma that Landau poles constitute fatal flaws of interacting quantum field theories.

\section*{Acknowledgments}

I would like to thank Tom DeGrand and Curtis Peterson for bringing the relevant lattice QED studies \cite{Kogut:1987cd,Gockeler:1997dn} to my attention, and W.~Ai and E.~Pr\'eau for helpful comments. Furthermore I would like to thank Bill Zajc for his help on putting Landau's stance on field theory in a historic context, Andrei Linde for very patiently explaining to me the wonderful work that has been done on this and related subjects in the 1970s \cite{Coleman:1974jh,Linde:1976qh,Abbott:1975bn,Kirzhnits:1978fy}, and Paul Stevenson for his objections to my section ``A Landau pole in QCD''.
This work was supported by the Department of Energy, DOE award No DE-SC0017905. 

\bibliography{PT}
\end{document}